# Ultralow Thermal Conductivity in a Two-Dimensional Material due to Surface-Enhanced Resonant Bonding

Sheng-Ying Yue[1], Tashi Xu[1,2], Bolin Liao[1*]

[1]Department of Mechanical Engineering, University of California, Santa Barbara, CA 93106, USA

[2]Department of Physics, University of California, Santa Barbara, CA 93106, USA

**Abstract**

Crystalline materials with ultralow thermal conductivity are highly desirable for thermoelectric applications. Many known crystalline materials with low thermal conductivity, including PbTe and $Bi_2Te_3$, possess a special kind of chemical bond called "resonant bond". Resonant bonds consist of superposition of degenerate bonding configurations that leads to structural instability, anomalous long-range interatomic interaction and soft optical phonons. These factors contribute to large lattice anharmonicity and strong phonon-phonon scattering, which result in low thermal conductivity. In this work, we use first-principles simulation to investigate the effect of resonant bonding in two dimensions (2D), where resonant bonds are in proximity to the surface. We find that the long-range interatomic interaction due to resonant bonding becomes more prominent in 2D due to reduced screening of the atomic-displacement-induced charge density distortion. To demonstrate this effect, we analyze the phonon properties of quasi-2D $Bi_2PbTe_4$ with an ultralow thermal conductivity of 0.74 W/mK at 300K. By comparing the interatomic force constants of quasi-2D $Bi_2PbTe_4$ and its bulk counterpart, and the properties of resonant bonds near the surface and in the bulk, we conclude that resonant bonds are significantly enhanced in reduced dimensions and are more effective in reducing the lattice thermal conductivity. Our results will provide new clues to searching for thermal insulators in low-dimensional materials.

**Keywords:** thermal conductivity, resonant bonding, 2D material, thermoelectrics

---

[*] To whom correspondence should be addressed. Electronic mail: bliao@ucsb.edu.

# Introduction

Crystalline materials with intrinsic low thermal conductivity are promising candidates for thermoelectric applications, as their electrical transport properties are less compromised by defects and disorders. Phonons, or quantized lattice vibrations, are the major heat carriers in crystalline semiconductors and insulators, and the thermal conductivity of these materials is determined by the specific heat, group velocity and scattering time of the phonon modes[1]. Intuitively, materials with heavy elements and weak bond strengths tend to have a lower thermal conductivity due to the reduced phonon group velocity[2,3]. It is less straightforward, however, to gain an intuition of the phonon-phonon scattering properties directly from the chemical composition and crystal structure of a material. A prominent example is boron arsenide, which, despite the heavy element arsenic, exhibits one of the highest thermal conductivities in inorganic materials due to its unique phonon-phonon scattering properties[4–7]. Therefore, direct connections between the chemical environment of constituent atoms and the phonon-phonon scattering properties of materials are valuable knowledge that can guide the search for materials with unusual thermal transport and thermoelectric properties[3].

One of such connections was recognized by Lee et al.[8] when they identified a type of chemical bonds, called "resonant bonds", shared by a group of known materials with low thermal conductivity[9], including IV-VI semiconducting compounds[10], $Bi_2Te_3$ and elemental Bi and Sb[11,12]. Resonant bonds form when equivalent configurations of chemical bonds exist given a specific crystal structure and number of bonding valence electrons[8,13]. The degeneracy, or "resonance", of the equivalent configurations results in a superposition of these configurations as the eventual bonding structure[13]. A simple one-dimensional example is illustrated in Fig. 1. If each atom possesses one valence electron in a one-dimensional atomic chain, there are two equivalent configurations for covalent bonds to form (configurations 1 and 2 in Fig. 1a). The resulting bonding structure is a superposition of the two configurations (Fig. 1b). This atomic chain is intrinsically unstable: an atomic displacement (Fig. 1c) will break the degeneracy and one bonding configuration will be preferred over the other, leading to a large distortion of the electronic charge density around the displaced atom (Fig. 1d). This distortion, partially electrostatically screened, will alter the electrostatic force field around neighboring atoms, break the bond degeneracy and cause propagation of the large charge density distortion. This "domino" effect leads to unusual long-range interactions between atoms along the resonant-bonded direction and destabilizes the one-dimensional chain. This electronic-degeneracy-induced instability is often referred to as Peierls instability[14] in condensed matter physics and Jahn-Teller effect in molecular and solid state chemistry[15]. Specifically, in IV-VI compounds such as PbTe, the valence *p*-electrons form highly directed networks of resonant bonds that lead to long-range interatomic interactions along particular directions, which in turn causes softened transverse optical (TO) phonons[8,16] and large phonon anharmonicity[17], both contributing to a low thermal conductivity.

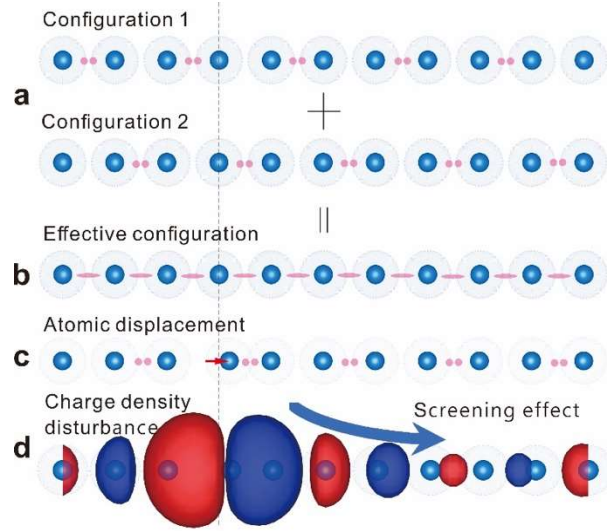

**Figure 1 Schematic for resonant bonds in a one-dimensional atomic chain.** (a) The two equivalent bonding configurations. (b) The effective bonding configuration as a superposition of the two equivalent configurations. (c) Atomic displacement breaks the degeneracy of the two equivalent configurations. (d) The spatial charge density disturbance caused by atomic displacement propagates along the chain after partial screening.

Since the propagation of large charge density distortion in a resonant-bonded network is limited by dielectric screening, it is expected that stronger long-range interactions will result from reduced screening in materials with resonant bonds. It is well known that dielectric screening in lower-dimensional materials is much weaker than that in three dimensions due to the geometric confinement of the induced charges responsible for screening[18]. For example, the electrostatic potential of a screened point charge in three dimensions decays exponentially in the form of Yukawa potential $\frac{e^{-r/r_0}}{r}$ ($r$ is the distance from the charge and $r_0$ is the screening length), whereas the electrostatic potential of a screened point charge in two dimensions decays logarithmically near the charge and asymptotes to the Coulomb form $\frac{1}{r}$ faraway from the charge[18]. In this light, we hypothesize that resonant-bonded networks in reduced dimensions will give rise to longer-ranged interatomic interaction than their 3D counterparts and thus lower thermal conductivity.

To verify this hypothesis, we use first-principles simulation to study phonon properties and thermal transport in quasi-2D $Bi_2PbTe_4$, whose crystal structure is presented in Fig. 2. The bulk form of this material (Fig. 2b) is predicted to be stable by Materials Project[19]. The bulk material comprises hexagonal-close-packing (hcp) septuple layers separated by van der Waals gaps, analogous to the quintuple layer structure of $Bi_2Te_3$[20]. Each septuple layer includes 4 layers of Te, two layers of Bi and one central layer of Pb. In this work, we investigate the phonon properties of one isolated septuple layer (Fig. 2a) and demonstrate that the surface-enhanced resonant bonding due to reduced screening leads to an ultralow thermal conductivity (0.76 W/mK at 300 K) compared to the bulk value of 1.44 W/mK at the same temperature. We reach this conclusion by detailed analysis of the distance dependence of the interatomic force constants and comparison of the resonant bonds near the surface and in the bulk.

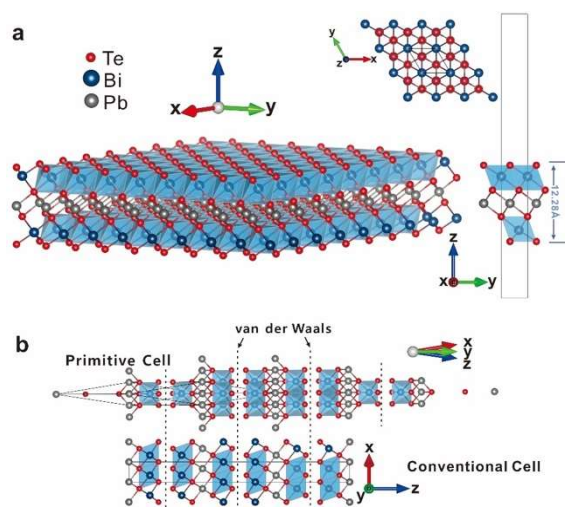

**Figure 2 Crystal structures of the quasi-2D and bulk $Bi_2PbTe_4$.** (a) The crystal structure of a septuple layer of $Bi_2PbTe_4$. Each septuple layer has a hexagonal close packing (hcp) structure, consisting of four Te layers, two Bi layers and one Pb layer. (b) The crystal structure, primitive cell and conventional cell of the bulk $Bi_2PbTe_4$.

## Results and Discussions

The details of the first-principles calculations are given in Methods. The calculated in-plane (isotropic within the x-y plane shown in Fig. 2a) thermal conductivity of both the bulk and quasi-2D $Bi_2PbTe_4$ within the temperature range of 200 K to 400 K is plotted in Fig. 3a. The thermal conductivity of the quasi-2D $Bi_2PbTe_4$ is significantly lower than that of the bulk. At 300 K, the thermal conductivity of the quasi-2D $Bi_2PbTe_4$ is 0.74 W/mK (compared to 1.44 W/mK of the bulk, which is much lower than other known 2D materials[21,22]. The significant difference between the quasi-2D and the bulk $Bi_2PbTe_4$ is unexpected given the identical internal crystal structure. We further compare the phonon dispersion relations of the bulk and the quasi-2D $Bi_2PbTe_4$ in Fig. 2b and c. Whereas the acoustic modes and high-frequency optical modes of the two systems are nearly identical, the soft transverse optical phonons ($TO_1$ and $TO_2$) have lower frequencies in the quasi-2D case. Furthermore, $TO_1$ and $TO_2$ modes in quasi-2D $Bi_2PbTe_4$ are significantly more anharmonic than those in bulk $Bi_2PbTe_4$, as reflected from the mode Grüneisen parameters shown in Fig. 3 (d) and (e). These low-frequency TO phonons are one feature of resonant-bonded materials and responsible for reduced thermal conductivity due to enlarged scattering phase space of acoustic phonons[8]. The lower frequency and higher anharmonicity of the TO modes in quasi-2D $Bi_2PbTe_4$ imply a more prominent effect of the resonant bonds in the quasi-2D system.

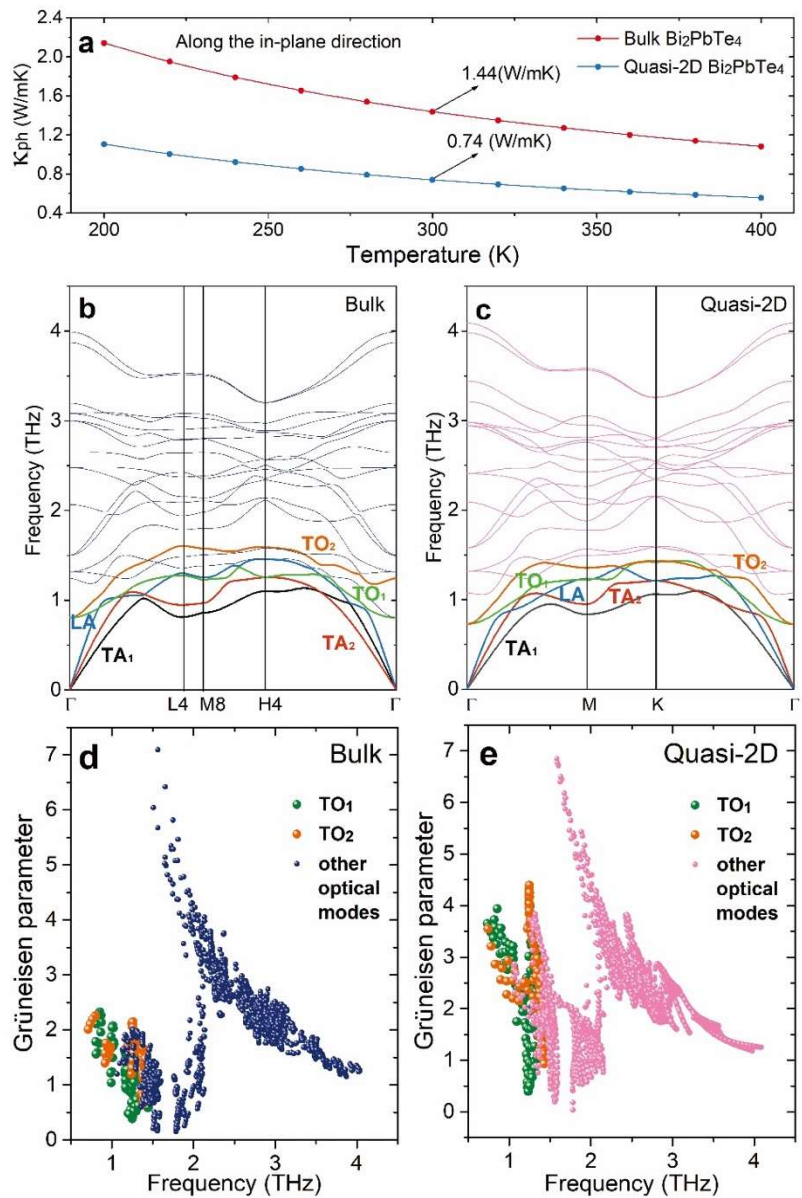

**Figure 3 Calculated lattice thermal conductivity, phonon dispersions and Grüneisen parameters of bulk and**

**quasi-2D Bi$_2$PbTe$_4$.** The lattice thermal conductivities of bulk and quasi-2D Bi$_2$PbTe$_4$ in (a) are both along the in-plane direction. The phonon dispersion relations of the two systems shown in (c) and (d) indicate lower frequencies of the TO modes in the quasi-2D system. The Grüneisen parameters shown in (d) and (e) reveal that the TO modes in quasi-2D Bi$_2$PbTe$_4$ are significantly more anharmonic than those in bulk Bi$_2$PbTe$_4$.

To further understand the origin of the discrepancy between the thermal conductivities of the bulk and the quasi-2D Bi$_2$PbTe$_4$, we calculate the second order interatomic force constants (IFCs) as a function of the distance between the corresponding atoms, as shown in Fig. 4 (the traces of the second order IFC tensors are normalized to those of the self-interacting IFC tensors and plotted here, as similarly done by Lee et al.[8]). In both the bulk (Fig. 4a) and the quasi-2D (Fig. 4b) Bi$_2$PbTe$_4$, we observe an anomalous nonmonotonic trend of the IFCs: in contrast to the monotonic decay of interaction strength versus distance as expected for conventional covalent or ionic bonds, large long-range interactions are present at specific neighboring shells, e.g. the fourth nearest neighbor. These large long-range IFCs are the consequence of directed resonant-bonded networks (Fig. 1). The fourth neighbor atom happens to be on the path of the resonant-bonded network, whereas the second and third neighbor atoms are not. This results in a much larger IFC between the fourth neighbor atom and the origin atom despite its longer distance from the origin atom than the second and third neighbor atoms. The unusual long-ranged interaction caused by resonant bonding also emerges for other neighbor shells, even inducing positive IFCs ("negative springs"), e.g. in 6th and 8th neighbor shells. Figure 4c-e show the charge density distortion caused by a small displacement of a central atom and felt by another atom, corresponding to the 4th-nearest-neighbor pairs labeled in Fig. 4b. These large charge density distortions caused by the breaking of degenerate bonding configurations (Fig. 1) lead to the strong long-ranged interatomic interactions along the bonding directions. More importantly, comparing Fig. 4a and 4b, we find that the strong resonant-bond-induced interactions decay slower and persist for a longer distance in the quasi-2D Bi$_2$PbTe$_4$ than those in the bulk. This indicates that the reduced dimensionality has a significant impact on the effective range of the resonant-bond-induced interatomic interaction, which, we hypothesize, can be explained by the different dielectric screening strengths in 3D and 2D.

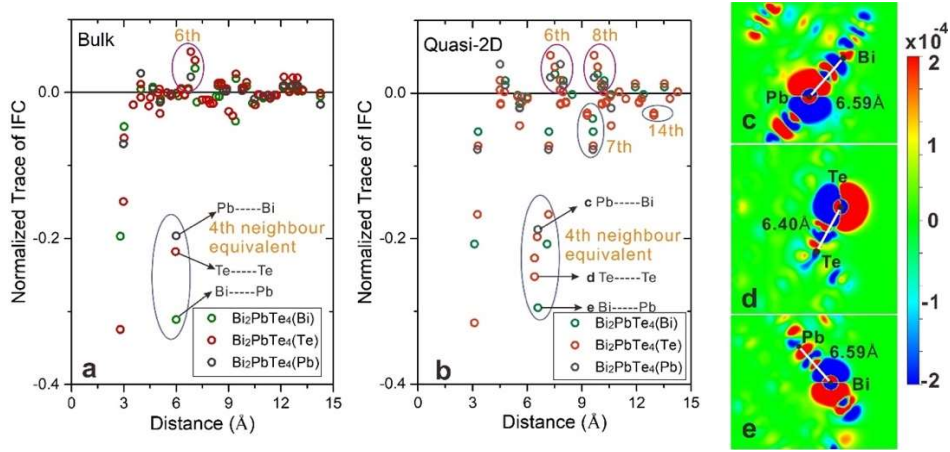

**Figure 4 Normalized traces of IFC tensors versus atomic distances and charge density disturbance of selected atom pairs in quasi-2D Bi$_2$PbTe$_4$.** Normalized traces of IFCs of (a) bulk Bi$_2$PbTe$_4$ and (b) quasi-2D Bi$_2$PbTe$_4$, where different colors denote IFCs with a specific element chosen as the origin atom, as indicated in the legend. (c-e) The charge density distortion in space due to the displacement of the origin atoms, corresponding to the selected atomic pairs labeled in (b). The color bar denotes the electron density disturbance with a unit of electron charge per Å$^3$.

To validate our hypothesis, we directly compare two Te layers in quasi-2D Bi$_2$PbTe$_4$ on the surface and inside the septuple layer, as illustrated in Fig. 5a and b. To contrast the strength and the range of the resonant bonds formed by Te atoms within the two layers, we first calculate the charge density distribution at equilibrium and compare the degree of electron delocalization in the two layers, as shown in Fig. 5c and d. Resonant bonds are known to create high charge density at the atomic intervals, i.e. a large degree of electron delocalization[8,23]. From Fig. 5c and d, it is clear that the electron density of the Te atoms on the surface is more delocalized, spreading more into the intervals among Te atoms, in comparison to that of the Te atoms inside the septuple layer, supporting our hypothesis that the resonant bonds formed near the surface are enhanced due to reduced screening. We further compare the charge density distortion caused by displacements of a Te atom in the two layers, as shown in Fig. 5e-f. The stronger and longer-ranged charge density distortions on the surface (Fig. 5e) as compared to those inside the septuple layer (Fig. 5f) provide direct evidence of surface-enhanced resonant bonds.

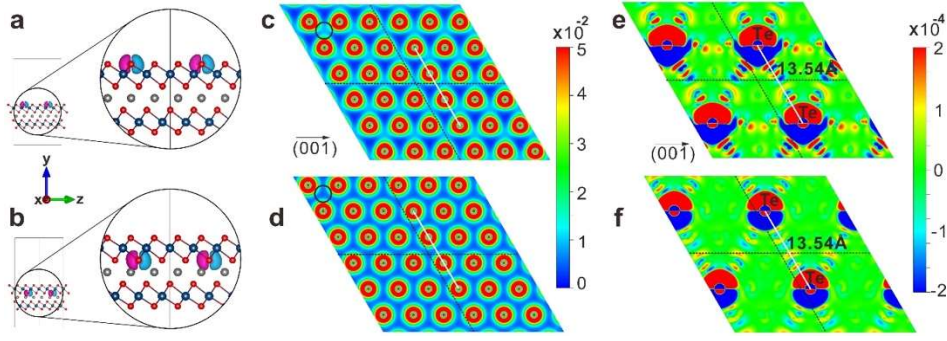

**Figure 5 Comparison of electron density distribution and displacement-induced charge density distortion on the surface and inside the septuple layer of quasi-2D Bi$_2$PbTe$_4$.** (a) and (b) indicate the Te layers on the surface and within the septuple layer that were investigated. (c) and (d) present the equilibrium electron density distribution of the surface and the internal Te atom layers, respectively. The color bar denotes the electron charge density in space in the unit of electron charge per Å$^3$. The black circles in (c) and (d) point to the interatomic region where a significant difference of the electron delocalization was observed between the surface and the internal layers. (e) and (f) compare the charge density distortion induced by the displacement of the Te atoms on the surface and within the septuple layer. The color bar denotes the electron density disturbance with a unit of electron charge per Å$^3$.

To quantify the different dielectric screening behaviors in the bulk and the quasi-2D Bi$_2$PbTe$_4$, we select the normalized traces of IFC tensors along resonant-bonded paths with the same Te atom chosen as the origin in both systems. The absolute values of these selected resonant-bond-enhanced IFCs are plotted as a function of interatomic distance in Fig. 6. To extract the effective screening lengths in the two cases, we use the Yukawa potential ($\frac{e^{-r/r_0}}{r}$) to fit the decaying trends, also shown in Fig. 6. We note here that Yukawa potential is associated with a screened point charge, as opposed to the delocalized charge density distortion in this case. Nevertheless, the Yukawa potential should agree well with the long-range data and provides a convenient measure, i.e. the screening length, to quantify the different screening strengths. The red and the blue solid lines are the best fits for the IFCs in the bulk and the quasi-2D Bi$_2$PbTe$_4$, respectively, and the yellow and the blue areas represent the ranges corresponding to $\pm 30\%$ change of the fitted screening length. From the fitting, we obtained that the average screening length of the large charge density distortion caused by resonant bonding in quasi-2D Bi$_2$PbTe$_4$ is 2.6 times longer that that in the bulk Bi$_2$PbTe$_4$. This result provides a quantitative measure of the surface enhancement of the resonant-bond-induced long-range interactions, which in turn lead to the significantly reduced thermal conductivity of the quasi-2D Bi$_2$PbTe$_4$ as compared to that of the bulk Bi$_2$PbTe$_4$.

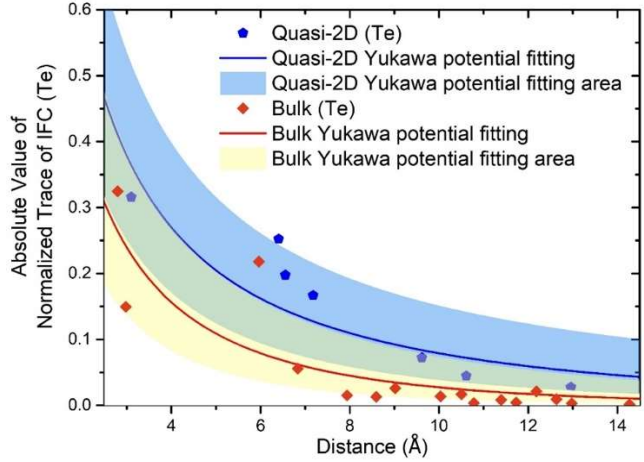

**Figure 6 Absolute values of the normalized traces of IFCs along resonant-bonded paths in the bulk and quasi-2D $Bi_2PbTe_4$ versus the interatomic distance.** The same Te atom are chosen as the origin atom in both cases. The decaying trends of the IFCs are fitted with the Yukawa potential with different effective screening lengths. The red and the blue solid lines are the best fits for the two cases, whereas the yellow and the blue areas correspond to $\pm 30\%$ change of the screening length from the best-fit values.

## Conclusion

In summary, we used first-principles calculation of a model material system ($Bi_2PbTe_4$) to investigate the effect of resonant bonds on thermal transport in reduced dimensions. We found that the thermal conductivity of quasi-2D $Bi_2PbTe_4$, i.e. one septuple layer of the material, is significantly lower than its bulk counterpart. By detailed analysis of the IFCs, charge density delocalization and the atomic-displacement-induced charge density distortion, we attributed the origin of the ultralow thermal conductivity of the quasi-2D $Bi_2PbTe_4$ to the surface enhancement of resonant bonds due to reduced dielectric screening. We further quantify the enhancement by fitting the interatomic distance dependent IFCs using the Yukawa potential, and we found the effective screening length in the quasi-2D $Bi_2PbTe_4$ is about 2.6 times longer than that in the bulk. The surface-enhanced resonant bonds lead to stronger and longer-range interatomic interactions and lower-frequency and more anharmonic TO phonon modes, which in turn contribute to the ultralow thermal conductivity. As the weakened dielectric screening is a general property of low dimensional materials, we envision that our finding of the surface-enhanced resonant bonds will serve as a general guiding principle to search for low dimensional materials with an ultralow thermal conductivity.

## Methods

We used the Vienna ab-initio simulation package (VASP)[24,25] for the density functional theory

(DFT) calculation in this work. The Perdew-Burke-Ernzerhof (PBE) generalized gradient approximation (GGA)[26] exchange-correlation functionals were used for the calculation. The pseudopotentials based on the projector augmented wave (PAW)[27] method were adopted. The cutoff of kinetic energy for wave functions was set at 820 eV (60 Ry)[8] and the energy convergence threshold was set as $10^{-8}$ eV. For the bulk and quasi-2D $Bi_2PbTe_4$, the Monkhorst-Pack[28] k-meshes of $6 \times 6 \times 6$ and $6 \times 6 \times 1$ were used to sample the Brillouin Zone (BZ). The convergence was checked here for the cutoff energy of the plane wave basis and the k-grid density. The vacuum spacing for the quasi-2D $Bi_2PbTe_4$ is larger than 20 Å to prohibit the atomic interaction across the van der Waals gap arising from the periodical boundary condition. The lattice parameters of both bulk and quasi-2D $Bi_2PbTe_4$ were optimized with the Hellmann–Feynman force tolerance $10^{-6}$ eV Å$^{-1}$. In addition, for the elements Bi, Te and Pb, the spin polarization calculations were included. The vdW-DF functionals were applied in VASP[29,30] for the bulk $Bi_2PbTe_4$ calculation to properly treat the interlayer van der Waals interactions.

**Lattice dynamics calculation**

We applied the density functional perturbation theory (DFPT) method[31] to calculate the lattice dynamics of the bulk and quasi-2D $Bi_2PbTe_4$. The harmonic second-order interatomic force constants (IFC) tensors were calculated by VASP combined with the PHONOPY package[32]. The IFC tensor is defined as:

$$\frac{\partial^2 E}{\partial R_i \partial R_j} = \begin{bmatrix} \frac{\partial^2 E}{\partial R_x \partial R_x} & \frac{\partial^2 E}{\partial R_x \partial R_y} & \frac{\partial^2 E}{\partial R_x \partial R_z} \\ \frac{\partial^2 E}{\partial R_y \partial R_x} & \frac{\partial^2 E}{\partial R_y \partial R_y} & \frac{\partial^2 E}{\partial R_y \partial R_z} \\ \frac{\partial^2 E}{\partial R_z \partial R_x} & \frac{\partial^2 E}{\partial R_z \partial R_y} & \frac{\partial^2 E}{\partial R_z \partial R_z} \end{bmatrix}, \quad (1)$$

where $E$ and $R_\alpha$ are the potential energy and atomic displacement along $\alpha$ direction. The trace of the IFC tensor reflects the bonding stiffness. To compare the IFCs in bulk and quasi-2D $Bi_2PbTe_4$ while excluding the impact of different bonding stiffness, we normalized the traces of IFC tensors between one atom chosen as the origin and another atom by the trace of the self-interaction IFC tensor of the origin atom, following the procedure by Lee et al.[8]:

$$\text{Normalized Trace of IFC} = \frac{\frac{\partial^2 E}{\partial R_{0,x} \partial R_{n,x}} + \frac{\partial^2 E}{\partial R_{0,y} \partial R_{n,y}} + \frac{\partial^2 E}{\partial R_{0,z} \partial R_{n,z}}}{\frac{\partial^2 E}{\partial R_{0,x} \partial R_{0,x}} + \frac{\partial^2 E}{\partial R_{0,y} \partial R_{0,y}} + \frac{\partial^2 E}{\partial R_{0,z} \partial R_{0,z}}}, \quad (2)$$

where $R_0$ denotes the displacement of the origin atom and $R_n$ the displacement of the n-th neighbor atom. The normalized traces of IFC for each atom in the bulk and quasi-2D $Bi_2PbTe_4$ are presented in Fig. 4a and b.

**Charge density distortion**

By applying the DFT calculations, we calculated the charge density distribution and its distortion as a response to small atomic displacements of the origin atoms for the quasi-2D $Bi_2PbTe_4$, which are presented in the Fig. 4 (c-e) and Fig. 5 (c-f). Figure 4(c-e) and 5(e, f) are the electron density distortion with the small atomic displacement and Fig. 5(c, d) are the equilibrium charge density distribution. The atomic displacement length is 2% of the distance between the center atom and the fourth neighbor atoms[8]. Because of the periodical boundary conditions, the resulting charge density distortion also has contributions from the periodic images of the central atom. To avoid this effect, we adopted the $2 \times 2 \times 1$ supercell (including 28 atoms) of the quasi-2D $Bi_2PbTe_4$ for the calculations. We obtained the charge density distortion by taking the difference between the electron density of the original and the displaced-atom cases[8].

**Lattice thermal conductivity**

Based on the harmonic IFC from the lattice dynamic calculations above, we obtained the phonon dispersion relations for the bulk and the quasi-2D $Bi_2PbTe_4$ via the PHONOPY package[32], which are shown in Figure 3 (b) and (c). In addition, through the changing volume method within the PHONOPY package[32], we also calculated the mode Grüneisen parameters using the definition $\gamma = -(V\,d\omega)/(\omega\,dV)$ of the bulk and the quasi-2D $Bi_2PbTe_4$, which are exhibited in Fig. 3(d) and (e). With this definition, the Grüneisen parameter ($\gamma$) evaluates the change of the frequency of phonon modes as a response to the material volume change, and thus larger $\gamma$ means higher anharmonicity of the corresponding phonon modes[33]. As the long-range interaction is significant in resonant-bonded materials, we adopted fine q-grid meshes to capture the long-range interaction in the bulk ($12 \times 12 \times 12$) and the quasi-2D $Bi_2PbTe_4$ ($12 \times 12 \times 1$).

To evaluate the lattice thermal conductivity, we also calculated the third-order (anharmonic) IFCs using a supercell frozen-phonon approach. $3 \times 3 \times 2$ and $3 \times 3 \times 1$ supercells were used here for the bulk and the quasi-2D $Bi_2PbTe_4$ calculations. Then the intrinsic lattice thermal conductivity ($\kappa_L$) was obtained by solving the phonon Boltzmann transport equation (BTE) iteratively as implemented in the ShengBTE package[34]. To include the long-range interactions, given our computing resources the cutoff radius ($r_{cutoff}$) was taken as $7.5\,\text{Å}$, which means that the interatomic interactions were considered up to the 6th nearest neighbors. In Fig. 4a and b, we see that the largest interatomic interactions are within 6th nearest neighbors for both bulk and quasi-2D $Bi_2PbTe_4$. In addition, we note that in the quasi-2D $Bi_2PbTe_4$, the 7th and 8th neighbor interactions are also significant as compared to the 6th neighbor interaction, implying that if we include all the long-range interactions in the quasi-2D $Bi_2PbTe_4$ the calculated thermal conductivity should be even smaller than the value 0.74 W/mK we reported here. The Born effective charges together with the dielectric tensor were also calculated and added to account for the long-range electrostatic interactions. For the quasi-2D $Bi_2PbTe_4$ thermal conductivity calculation, the material thickness (13.48 Å) used was taken as the sum of the thickness of the real atomic layers and the van der Waals radii of the surface Te atoms[35]. We checked the convergence of $\kappa_L$ with the q-grid mesh size. For the bulk and the quasi-2D $Bi_2PbTe_4$, we adopted the q-grid meshes $12 \times 12 \times 12$ and $30 \times 30 \times 1$, respectively.

## Author contributions

S.-Y. Y. and B. L. conceived this project; S.-Y. Y. and T. X. carried the first-principle calculations. S.-Y. Y. and B. L. analyzed the data and wrote the paper. S.-Y. Y. and T. X. contributed equally to this work.

## Acknowledgements

This work is supported by a startup fund from UCSB. B. L. acknowledges the support provided by the Regents' Junior Faculty Fellowship and an Academic Senate Faculty Research Grant from UCSB. We acknowledge computational resource support from the Center for Scientific Computing from the CNSI, MRL: an NSF MRSEC (DMR-1720256) and NSF CNS-1725797. This work also used the Extreme Science and Engineering Discovery Environment (XSEDE)[36] Stampede 2 at the Texas Advanced Computing Center (TACC) through allocation TG-DMR180044. XSEDE is supported by NSF grant number ACI-1548562.